# Porous SiO$_2$ coated dielectric metasurface with consistent performance independent of environmental conditions

**RENÉ GEROMEL[1,*], CHRISTIAN WEINBERGER[2], KATJA BRORMANN[2], MICHAEL TIEMANN[2] AND THOMAS ZENTGRAF[1]**

[1] *Department of Physics, Paderborn University, Warburger Straße 100, D-33098 Paderborn, Germany*

[2] *Department of Chemistry, Paderborn University, Warburger Straße 100, D-33098 Paderborn, Germany*

[*] *rgeromel@mail.upb.de*

**Abstract:** With the rapid advances of functional dielectric metasurfaces and their integration on on-chip nanophotonic devices, the necessity of metasurfaces working in different environments, especially in biological applications, arose. However, the metasurfaces performance is tied to the unit cell's efficiency and ultimately the surrounding environment it was designed for, thus reducing its applicability if exposed to altering refractive index media. Here, we report a method to increase a metasurface's versatility by covering the high-index metasurface with a low index porous SiO$_2$ film, protecting the metasurface from environmental changes while keeping the working efficiency unchanged. We show, that a covered metasurface retains its functionality even when exposed to fluidic environments.

## 1. Introduction

Metamaterials are artificially designed structures whose optical properties do not primarily arise from the choice of material itself but rather from the design and distribution of so called single meta-atoms. These meta-atoms are normally arranged in a sub-wavelength periodic lattice and allow for an unparalleled wavefront control near their resonance. As its 2D-variant, metasurfaces excel at compact applications in on-chip nanophotonic devices [1-3]. Related advanced research fields include linear [4-6] and nonlinear imaging [7,8] and holography [9-12], optical cloaking [13,14] and even biomedical applications such as the sensing of proteins [2,15] or the use of optical tweezers to confine and manipulate particles [16-20].

Especially all-dielectric metasurfaces benefit from applications that work in transmission and their efficiency is determined by the single unit cell that consists of high-index nanostructures in a low-index environment (generally air) [21,22]. Because the designed utility will usually be achieved in a predetermined environment the properties of the metasurface change when the refractive index of the surrounding material is changing. Furthermore, in order for the metasurfaces to work at high diffraction efficiencies, it is beneficial to choose a high refractive index contrast between the nanostructures and the surrounding medium, allowing for tighter confinement of the excited modes and a reduced height of the structures.

As a consequence, most meta-atom designs are rendered suboptimal if transferred to another refractive index environment and would need to be redesigned to work properly. To illustrate this effect, we calculated the transmission of a typically geometric-phase metasurface made of silicon nanofins on a glass substrate for different refractive index values of the surrounding host material $n_H$. The effect of a varying refractive index on the transmission behavior of a unit cell can be seen in Figure 1a. For geometric-phase metasurfaces, phase information is applied to the transmitted light via polarization conversion of an incident circularly polarized state.

Therefore, we calculated the transmission for both circular states whereas co-polarization signifies the same helicity as the incident light and cross-polarization represents the converted polarization that carries the desired phase information. Both the transmitted co- and cross-polarization change significantly with the refractive index of the environment thus reducing the metasurface's efficiency, e.g., the ability to alter the phase of the transmitted light in geometric-phase metasurfaces (see Fig. 1a).

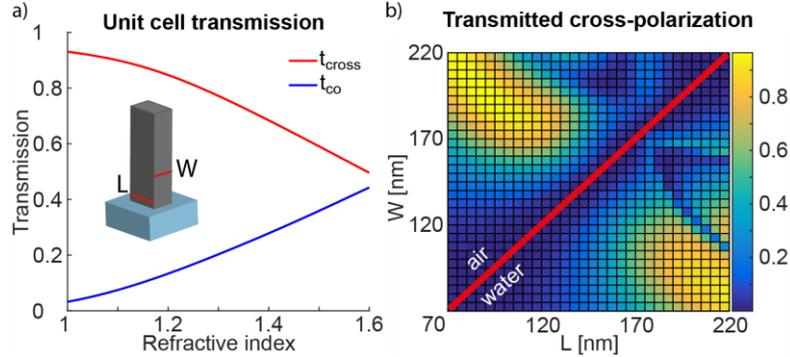

Fig. 1: a) Transmission for the co- and cross-polarization of a metasurface made of high index Si nanofins (length L = 180 nm, width W = 95 nm, height H = 600 nm) in dependence of the host medium's refractive index $n_H$. An increase in the refractive index reduces the conversion efficiency of the unit cell (shown in the inset). b) Calculated transmission of the cross-polarization for a variation in Si nanofin length and width. The values above the red line are plotted for the transmission with air as the host medium whereas the values below the red line represent the transmitted cross-polarization in the case that water surrounds the Si nanofins. The calculations of both panels were done using the RCWA method. The simulation is done for a wavelength of 800 nm, a Si height of 600 nm and a unit cell periodicity of 400 nm.

## 2. Metasurface design and fabrication

By changing the surrounding environment, the host medium of the meta-atoms also changes, thus typically resulting in a reduced transmission and diffraction efficiency. In this work, we provide a solution to this problem by coating the metasurface with porous $SiO_2$, hence preventing any changes to the refractive index of the surrounding medium that is in direct contact with the meta-atoms by creating a buffer layer. As noted before, the highest efficiency for dielectric metasurfaces can be obtained for high-index structures in a low-index environment. Therefore, it is desired to embed the nanostructures into a material with a refractive index close to one. As for the design of the metasurface itself, we utilize the Panacharatnam-Berry phase concept for metasurfaces [23] and optimize the cross-polarization transmission for a wavelength of 800 nm. Figure 1b shows the results for varying parameters of the width and length of the Si nanofin and different host media (air $n_H$=1 and water $n_H$=1.3). The initial parameter sets are chosen as 200 nm x 80 nm with a height of 600 nm to give a high transmission of nearly 97% if the structures are embedded in air. From Figure 1b, one can see that the transmission for these sets of parameters already drops by 10% when the structures are embedded in water.

Next, we fabricate the metasurfaces for these parameters by patterning an amorphous silicon film via electron beam lithography and dry etching. The fabrication follows a standard procedure (Figure 2a), by depositing 600 nm amorphous silicon on a glass substrate and spin-coating 270 nm of positive PMMA photoresist and 35 nm of Electra 92 as a conducting layer. After patterning via electron beam lithography with an acceleration voltage of 20 kV at a dose of 220 μC/cm², and developing the exposed PMMA resist, a 20 nm thick chromium mask is deposited as an etching mask for the silicon. The following reactive ion etching utilizes a pseudo-Bosch process using $SF_6$ as the etching gas and $C_4F_8$ as the polymerization gas. Finally, the chromium mask is removed using chromium etch and the porous $SiO_2$ is applied via a dip-coating procedure.

The synthesis of these films is based on the work of Lozano-Castelló [24]. Tetraethylorthosilicate, Pluronic F-127, hydrochloric acid, water, and ethanol were mixed in a molar ratio of 1: $6.6 \cdot 10^{-3}$: $6.6 \cdot 10^{-3}$: 4.62: 22.60. After plasma cleaning for 20 min the substrate was dip-coated with the above-mentioned solution at a relative humidity of 20% and a constant speed of 4 mm s$^{-1}$. After curing of the film at 120 °C for about 1 h the sample was calcined at 450 °C (1 °C min$^{-1}$) for 5 h in air. The coating procedure was repeated three times, leading to a 1 µm thick highly porous film and thus to a minimal refractive index.

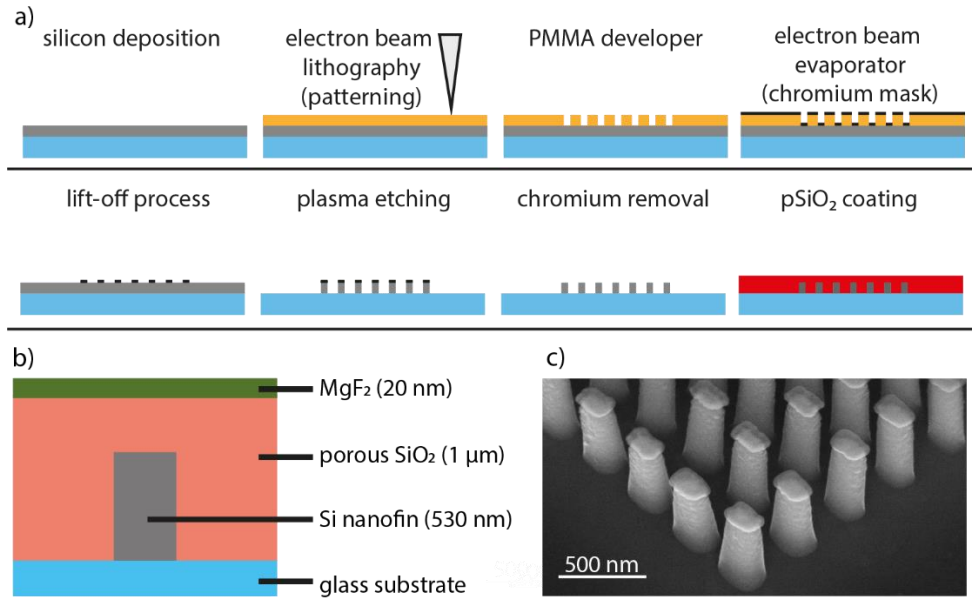

Fig. 2: a) Fabrication procedure for the metasurfaces. Following silicon deposition, the metasurfaces are patterned on a positive PMMA photoresist using electron beam lithography. After the PMMA development, a chromium mask is used as a protective mask for the ensuing etching process. Finally, the chromium mask is removed and the porous $SiO_2$ film is applied via dip-coating. b) Unit cell cross-section of the finished sample. As the final step, an $MgF_2$ layer is added to close the pores of the sample. c) SEM image of a fabricated sample before the chromium mask removal under a viewing angle of 45°.

To prevent other materials like liquids from penetrating the pores of the porous $SiO_2$, we apply an additional film of a transparent dielectric material. This step is supposed to protect the pores from filling for example with water (see Section 3). Here, we use a 20 nm thick $MgF_2$ layer deposited by electron beam evaporation. The final sample layout is shown in Figure 2b. Because of the large distance between the Si nanofins and the $MgF_2$ layer, no influence on the optical properties is expected due to the localized modes inside the silicon nanofins.

After producing the mentioned design, a few iterative adjustments have been made to the nanofin dimensions to improve the measured transmissive spectra that deviated from the simulations due to tolerances in the fabrication and some uncertainties in the refractive index value of the amorphous silicon. In the eventually used design, the height was lowered to 530 nm while the length and width were fixed at 240 nm x 120 nm. A scanning electron microscopy image of one sample made before the chromium mask removal is shown in Figure 2c.

## 3. Porous SiO₂ film characterization

For the use in optical applications, the coating of the metasurface has to fulfill certain criteria: First, the refractive index of the coating should be as low as possible because a high refractive index lowers the maximum conversion efficiency of the Si nanofins as they will not allow for tight confinement of the excited modes. Second, the surface roughness should be small as a rough surface can alter the transmitted wavefronts and impair the metasurface's functionality. With both efficiency and functionality suffering from an improper coating its applicability would be severely limited and therefore investigating the refractive index as well as the surface roughness is of great interest when concluding the viability of this particular coating.

Here, the porous $SiO_2$ film was characterized via ellipsometry and atomic force microscopy (AFM) to measure the refractive index and surface roughness corresponding to the used recipe mentioned in Section 2. The refractive index measurement is shown in Figure 3a. In the spectral range from 500 nm - 900 nm, the refractive index decreases gradually towards longer wavelengths and reaches 1.182 at the target wavelength of 800 nm, lower than related work from Wang et al. [25] of 1.23 who also employed a sol-gel approach. Work done by Xi et al. [26] even reports a film deposition with a refractive index of as low as 1.05 using electron beam evaporation of $SiO_2$ granules at low vapor flux and deposition under 85° substrate tilt angle resulting in the formation of tilted $SiO_2$ nanopillars. While these coatings are very beneficial for antireflection coatings, their applicability as a protective layer for metasurfaces might be unsuited especially since the deposition method will introduce shadowing effects due to the already present Si nanofins potentially resulting in an inhomogeneous coating.

Regarding the exposition of our samples to water, the influence of varying humidity in the environment on the refractive index is shown in Figure 3b. The measured refractive index rises with increasing humidity indicating that water is penetrating the pores which will impact the performance of the coated samples. As mentioned in Section 2, an additional $MgF_2$ film is supposed to prevent this by closing the top pores and waterproofing the sample. The porous film also exhibits a small hysteresis, showing that the water does not immediately leave the film once the surrounding humidity is lowered again.

As for the AFM measurement, the evaluated surface roughness yields a root mean square (RMS) value of 3.5 nm. Regarding the typically used wavelengths ranging from visible to infrared for metasurface applications, the surface of the porous $SiO_2$ layer can be considered flat which is a desirable property for optical applications.

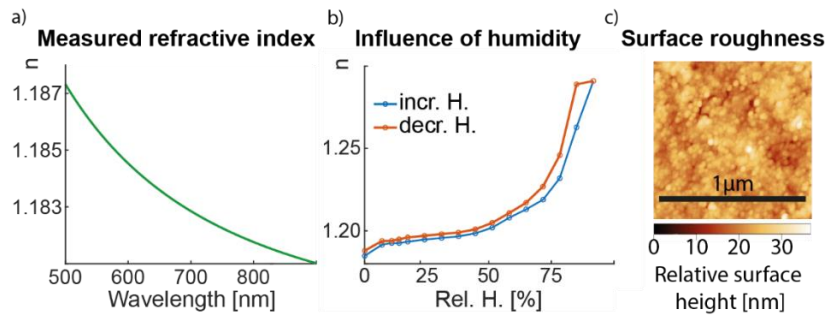

Fig. 3: a) Ellipsometry measurement of a porous $SiO_2$ layer for a wavelength range of 500 nm - 900 nm. The refractive index (n) decreases towards longer wavelengths and reaches a value of 1.182 at 800 nm. b) Measured refractive index (n) of a porous $SiO_2$ layer for increasing and decreasing relative humidity (Rel. H.). The refractive index increases with higher humidity; liquid environments, e.g. water, can enter the pores and thus change the host medium's refractive index. The porous $SiO_2$ layer also exhibits a hysteresis for decreasing humidity which means that once exposed to water the humidity partly remains in the layer and does not immediately leave the pores. c) AFM measurement of the porous $SiO_2$ layer's surface. The surface roughness yields an RMS value of 3.5 nm.

## 4. Sample characterization and optical measurements

To evaluate the influence of the porous $SiO_2$ coating on the meta-atoms' performance, first, the transmitted co- and cross-polarization intensities ($I_{co}$ and $I_{cross}$), as well as the conversion efficiency $\eta_{conv} = I_{cross}/(I_{cross} + I_{co})$ of the metasurface before and after the coating via porous $SiO_2$ are determined. Since the metasurface is designed to act as a half-wave plate, $I_{cross}$ represents the intensity converted to its cross-polarization for circularly polarized light (opposite helicity). As for the metasurfaces, additional samples containing a linear phase gradient based on the geometric phase [23] have been fabricated to diffract the light under angles of 0°, 5°, and 10°. For the sample with the 0° diffraction angle where all the nanofins are oriented the same way, the transmission from 700 nm – 1000 nm was measured with a Bruker Hyperion FT-IR microscope. Albeit, only including linear polarizers, if these are oriented at ±45° towards the rectangular nanofins' principle axis the broadband transmission measurement effectively averages the two possible co- or cross-polarization measurements, namely L→L and R→R for the transmitted co-polarization and L→R as well as R→L for the cross-polarization. Figure 4 shows the transmitted co- and cross-polarization intensities and the conversion efficiency with and without the porous $SiO_2$ coating. Before applying the coating, the sample shows cross-polarization peaks around 825 nm and 940 nm whereas the conversion efficiency shows two peaks at 845 nm and 940 nm reaching up to 95%. After the coating, the peaks shift by roughly 40 nm towards longer wavelengths. The conversion efficiency also decreases by up to 5% at the peak positions. While a small decrease in efficiency is expected, note that the fabrication parameters were optimized for air as the host medium. For proper operation at maximum efficiency, the unit cell's parameters need to be calculated considering the porous coating as the surrounding refractive medium.

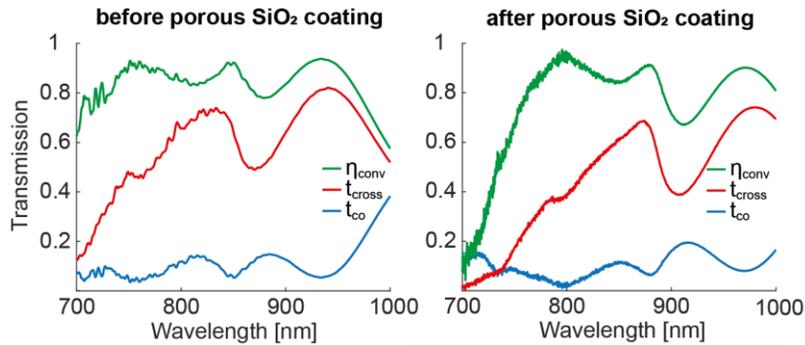

Fig. 4: Broadband transmission measurement of co- and cross-polarization ($t_{co}$ and $t_{cross}$) as well as conversion efficiency ($\eta_{conv}$) for the 0° diffraction angle metasurface. (left) Spectra for the fabricated sample before coating with porous $SiO_2$. (right) Spectra after applying a 1 µm thick porous $SiO_2$ layer. After the coating, the peaks shift by roughly 40 nm towards longer wavelengths.

The results of the transmitted cross-polarization from Figure 4 (right) will be the baseline for comparison to evaluate the following measurements. After depositing a 20 nm layer of $MgF_2$ the measurements using the FT-IR microscope were repeated (see Figure 5 (left)). To expose the metasurface to water, a droplet has been placed on top of the fields. The blue curve shows that although the spectrum keeps a similar shape, the derivation from the original curve is neither negligible nor satisfactory with the transmission decreasing by as much as 20% near 1000 nm. The measurement implies that despite the supposed protection from the $MgF_2$ layer water penetrated into the porous $SiO_2$ film and reached the nanofins. This might be due to the $MgF_2$ layer also being porous when deposited via electron beam evaporation. The change in spectral behavior however is less pronounced compared to the difference from Figure 4 since the spectrum shifts by only 20 nm towards longer wavelengths. Still, to resolve the issue of water penetrating the porous film an additional layer of 500 nm PMMA is applied to the surface via spin-coating. The film thickness of 500 nm was chosen to ensure that we obtain a dense and closed layer even covering the edges of the sample substrate. In hindsight, applying the waterproofing layer via dip-coating might be more reliable than a spin-coating approach because it can cover the uneven edges better. The measurements show that the water now barely influences the spectral behavior of the sample (Figure 5). Note that the placed water droplet on the sample surface slowly evaporated during the measurements and that the focus of the FT-IR microscope had to be adjusted repeatedly, potentially leading to less light collection from the condenser lens which could explain the slightly lower transmission values. Hence, small deviations in the spectra might arise from the re-adjustment of the setup and some inhomogenous thickness of the water layer. However, the results show that the porous $SiO_2$ coating in combination with a waterproofing layer allows the metasurface to retain its efficiency and specral characteristics if exposed to different refractive media.

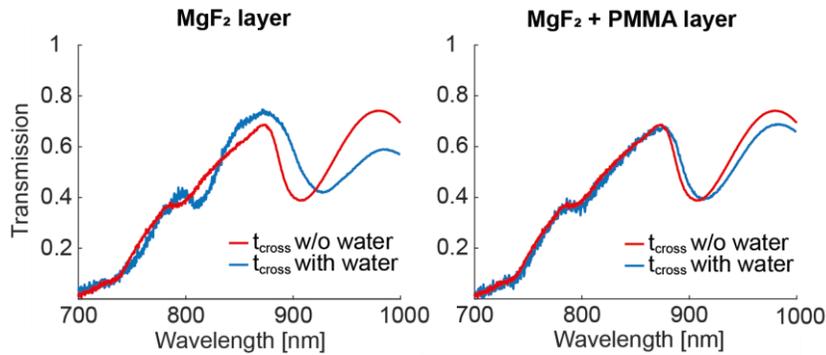

Fig. 5: (left) Comparison between the transmitted cross-polarization after applying the porous $SiO_2$ coating (see also Fig. 4 right) and the transmitted cross-polarization after adding a 20 nm $MgF_2$ layer as well as exposing the metasurface to water. The transmission for these two curves differs noticeably. The measurement indicates that the water entered the porous $SiO_2$ layer despite the $MgF_2$ film and thus changed the refractive index of the host medium. (right) Here, an additional 500 nm PMMA film has been added to realize a waterproofing layer. These two curves show a close resemblance and the water barely seems to influence the metasurface's transmission.

In addition to the previous measurements, we simulated the spectral behavior of these metasurfaces when exposed to different environments. The left panel of Figure 6 shows the simulated transmitted cross-polarization for an uncoated sample exposed to air and water whereas the right panel of Figure 6 shows the influence of the porous coating and compares the simulated transmitted cross-polarization if a coated Si antenna is exposed to air and water. Note that the simulation does not include the additionally applied $MgF_2$ and PMMA layers. Focusing on the peak shown at 830 nm, it can be seen from Figure 6 (left) that the transmitted cross-polarization and thus the overall efficiency of the metasurface drops due to the changed environment. Also, the dip for the uncoated metasurface located at 870 nm shifts by roughly 40 nm towards longer wavelengths when exposing it to water. If the metasurface is embedded in the porous $SiO_2$ we only observe a spectral shift of 20 nm (Figure 6 (right)). More importantly, after coating there is negligible influence in the spectral behavior for air and water exposition (blue and green curves). The simulated curves mostly match with the measurements albeit showing an overall larger transmission than the experimentally measured ones. Note that the measured and simulated curves not only represent the spectral behavior of the 0° phase gradient metasurface but also of the 5° and 10° phase gradient metasurfaces since the behavior of the entire sample originates from the spectral behavior of the unit cell. However, small deviations might occur due to weak coupling between neighboring nanoantennas.

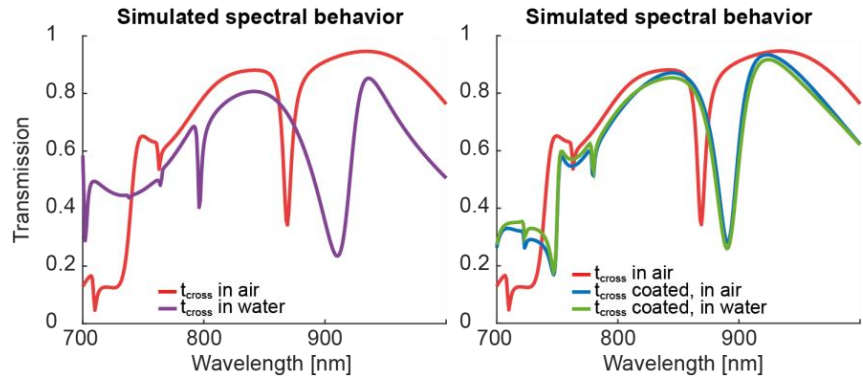

Fig. 6: (left) Simulated transmission of the cross-polarization state of the uncoated metasurface when exposed to air and water. A spectral shift to larger wavelengths by roughly 40 nm can be observed if the sample is exposed to water. (right) Simulated transmission of the cross-polarization state of the uncoated metasurface as well as the transmission of the porous $SiO_2$ coated sample when exposed to air and water. The difference between the latter two is marginal.

With the metasurface retaining its efficiency after the coating, the gradient-phase fields that diffract the light under an angle of 5° and 10° are used to investigate the functionality for the finished sample with and without water. Here, we use polarization optics to realize circularly polarized light and focus 800 nm laser light on the gradient fields. Depending on the circular input polarization (L or R) the +1 or -1 diffraction order will show up in the transmitted light. By imaging the Fourier plane of the collecting microscope objective, we spatially separate the k-vectors of the transmitted light, these being the +1 and -1 diffraction order (cross-polarization) as well as the light that did not interact with the metasurface (co-polarization). Figure 7a shows an illustration of the used 4f-setup and Figure 7b shows the obtained camera images from the measurement of the 5°- and 10°-diffraction angle gradient-phase metasurfaces for incident left and right circular polarized light, respectively. For the measurement with a fluidic environment, the sample was placed in a cuvette filled with water.

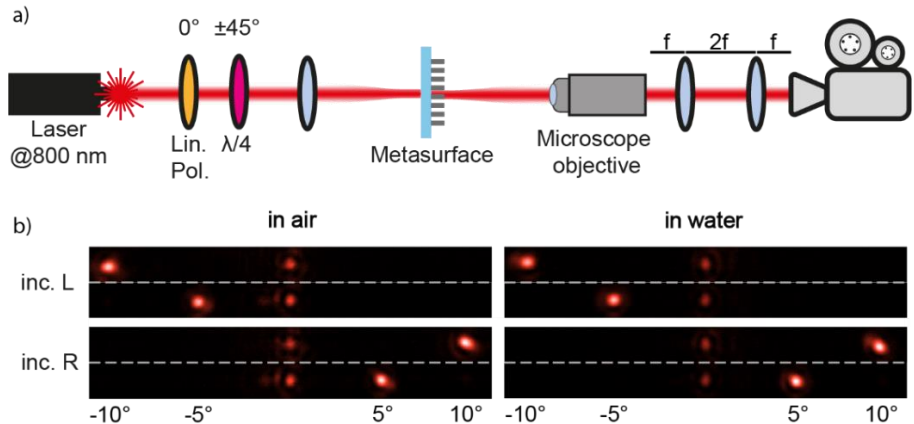

Fig. 7: (a) Illustration of the 4f measurement setup. A linear polarizer and a λ/4-plate are used to create a circularly polarized state and two lenses behind the microscope spatially separate the diffraction orders that are captured on a camera. (b) Camera images showing the diffraction orders of the 5° and 10° linear phase gradient metasurfaces in the Fourier plane at 800 nm. (left) +1 and -1 diffraction orders (incident L and R) for metasurfaces that are exposed to air. (right) +1 and -1 diffraction orders from the same metasurfaces after placing the sample in water. The diffraction orders (location and strength) do not noticeably change when exposing the sample to water.

The measurements show, that the diffraction angles and the strength of the diffraction spots stay the same even after placing the sample in water. The measurements on such geometric-phase metasurfaces illustrate that the functionality stays unaffected if the sample is placed inside an environment with a different refractive index. The small tradeoff in slightly lower maximum diffraction efficiency due to the porous $SiO_2$ coating compared to air as the host medium can be considered beneficial if the metasurface is supposed to operate in an environment with varying refractive index. Furthermore, the porous $SiO_2$ coating also acts as a protection layer for the small nanostructures, making the sample surface more robust in harsh environments.

## 5. Conclusion

In this work, we present a method to protect a dielectric metasurface from the influence of a changing environment by coating the sample in a porous $SiO_2$-film. The low refractive index of this film allows the metasurface to keep a relatively large index contrast between the dielectric nanostructures and the surrounding host medium. By fixing the nanostructures' host medium, especially to one with a low refractive index, the metasurface retains its functionality at high efficiency and can even be placed in fluidic environments without changing the functionality. However, since the porous coating can easily soak itself with liquids like water and thus change the index contrast between the host medium and the meta-atoms, waterproofing by adding another transparent layer is necessary. To solve this, we applied a layer of PMMA via spin-coating. For this procedure to work, a sufficiently smooth sample surface or a thick PMMA film is required to ensure the waterproofing. Employing another dip-coating procedure, similar to how the porous $SiO_2$-film was applied, might proof to be more reliable, especially regarding the edges of the sample substrate that will not be covered by spin-coating alone. Furthermore, the porous coating also acts as a protection against physical damages and might also overcome one of the hurdles for commercializing the incorporation of metasurfaces in everyday devices, as suggested by Gun-Yeal Lee et al. [11].

**Acknowledgment.** The authors like to thank Cedrik Meier for providing access to the EBL system and Basudeb Sain for help with the sample fabrication.

**Funding.** This project has received funding from the European Research Council (ERC) under the European Union's Horizon 2020 research and innovation programme (grant agreement No 724306) and the Deutsche Forschungsgemeinschaft (DFG No. ZE953/11-1).

**Disclosures.** The authors declare no conflicts of interest.

**Data availability.** Data underlying the results presented in this paper are not publicly available at this time but may be obtained from the authors upon reasonable request.